\def\rfr#1{eq. (\ref{#1})}
\def\rfrs#1#2{eqs. (\ref{#1})-(\ref{#2})}

\def\Rfr#1{Eq. (\ref{#1})}

\def\dert#1#2{\frac{{{d}}{#1}}{{{d}}{#2}}}              % derivate parziali e totali prima e seconda

\def\asec{$''$ cy$^{-1}$}

\def\bar{\begin{eqnarray}}
\def\ear{\end{eqnarray}}
\def\bb{\bibitem}
\def\eqi{\begin{equation}}
\def\eqf{\end{equation}}
\def\eqia{\begin{eqnarray}}
\def\eqfa{\end{eqnarray}}
\def\rp#1#2{{\frac{#1}{#2}}}
\def\ct#1{\cite{#1}}
\def\lb#1{\label{#1}}

\documentclass[11pt]{article}
\usepackage{amsmath,amsthm,amscd,amssymb}
\usepackage{latexsym}
\usepackage{graphicx,epsfig}
\makeindex \textwidth 160mm \textheight 230mm \topmargin -5mm
\begin{document}

\noindent{\bf \LARGE{What do the orbital motions of the outer
planets of the Solar System tell us about the Pioneer anomaly?}}
\\
\\
\\
{Lorenzo Iorio}\\
{\it Viale Unit$\grave{a}$ di Italia 68, 70125\\Bari, Italy
\\e-mail: lorenzo.iorio@libero.it}\\\\
{Giuseppe Giudice}\\
{\it Dipartimento di Progettazione e Gestione Industriale,\\
Piazzale Tecchio 80, 80125
\\Napoli, Italy
\\e-mail: giudice@unina.it}

\begin{abstract}
In this paper we investigate the effects that an anomalous
acceleration as that experienced by the Pioneer spacecraft after
they passed the 20 AU threshold would induce on the orbital
motions of the Solar System planets placed at heliocentric
distances of 20 AU or larger as Uranus, Neptune and Pluto. It
turns out that such an acceleration, with a magnitude of
$8.74\times 10^{-10}$ m s$^{-2}$, would affect their orbits with
secular and short-period signals large enough to be detected
according to the latest published results by E.V. Pitjeva, even by
considering errors up to 30 times larger than those released. The
absence of such anomalous signatures in the latest data  rules out
the possibility that in the region 20-40 AU of the Solar System an
anomalous force field inducing a constant and radial acceleration
with those characteristics affects the motion of the major
planets.
\end{abstract}


\section{Introduction}
The Pioneer 10/11 missions were the first spacecraft to explore
the outer Solar System. Pioneer 10 was launched in March 1972;
Pioneer 11 followed it in April 1973. After the encounters with
Jupiter and Saturn, both Pioneer spacecraft continued to explore
the outer regions of the Solar System following nearly ecliptical
hyperbolic paths in opposite directions. The last successful
communication with Pioneer 10 occurred in April 2002, at nearly 70
AU from our planet, while the last scientific observations were
returned by Pioneer 11 in September 1995 when it was at
approximately 40 AU from the Earth. The last attempt to
communicate with Pioneer 10 was made on March 2006, but without
success \cite{toth}. The so-called Pioneer anomaly consists in an
anomalous constant and uniform acceleration directed towards the
Sun of $(8.74\pm 1.33)\times 10^{-10}$ m s$^{-2}$ found in the
data of both spacecraft from the moment when they passed the
threshold of 20 AU \cite{And98, And02, Tur05}. Various attempts to
explain this feature in terms of known physical and engineering
effects have been done, but without success; the nature of the
Pioneer anomaly is, at present, still unexplained. It is not clear
if the anomaly is due to some internal systematic, as suggested,
e.g., in \cite{Kaz}, or has an external origin. Dedicated space
missions have recently been proposed \cite{Tur04, Chui}, while the
authors of \cite{Izzo06} put forth two non-dedicated scenarios
involving either a planetary exploration mission to the outer
Solar System or a piggy-backed micro satellite to be launched
toward Saturn or Jupiter.

In this paper we propose to analyze some aspects of the problem of
the Pioneer anomaly. In particular, we will investigate the
possibility that an external, unknown constant and uniform force
field inducing an acceleration of $8.74\times 10^{-10}$ m s$^{-2}$
on a test particle is present in the outer regions of the Solar
System within 20-40 AU. The outline of our approach is the
following.

Let us consider a constant and uniform acceleration radially
directed towards the Sun and of very small magnitude so that it
can be treated with the usual perturbative approaches; we will
deal with it from a purely phenomenological point of view, without
making any speculation about its origin. For a review of many
proposed mechanisms involving possible `new' physics, see
\cite{dittus} and references therein. What is its impact on the
orbital motion of the planets of the Solar System whose semimajor
axis is larger than 20 AU like Uranus, Neptune and Pluto? it
should be noted that, at present there are no data supporting a
more gradual onset of such an anomalous effect, and in the
framework of this simplified model such planets, due to their
relatively small eccentricities ($e_{\rm Uranus}=0.047$, $e_{\rm
Neptune}=0.008$ and $e_{\rm Pluto}=0.248$), would be acted upon by
the anomalous acceleration in all portions of their orbits.

An analogous approach was recently followed by the authors of
\cite{Page06}. Contrary to us, they discard the possibility of
using the major planets of the outer Solar System.
%They write:
%``[...] the ephemerides of the outer planets are almost entirely
%based on optical observations \cite{Sta04} and are much less
%accurate than those for the inner planets. [...] Modern astrometry
%can obtain angular positions with reasonable accuracy, but the
%mean motions of the outer planets and their radial distances are
%quite uncertain. Thus, any Pioneer effect perturbation on these
%bodies would be masked by uncertainty in the orbital semimajor
%axis. Given these uncertainties, we must conclude that the outer
%planets do not represent good candidates for astrometrically
%determining the reality of the Pioneer Effect.''
The focus of
\cite{Page06} is on some selected minor bodies characterized by
large semimajor axes and eccentricities like some unusual minor
bodies, Trans-Neptunian Objects (TNOs) and Centaurs. However, such
objects are not constantly under the action of the hypothetical
anomalous acceleration under investigation because of their highly
eccentric orbits ($0.620<e<0.947$). According to the authors of
\cite{Page06}, a relatively low-cost future astrometric campaign
should resolve the problem within the next twenty years. A
different position about the outer planets can be found in
\cite{Izzo06}.
%In it it is written: ``[...] a modification of
%gravitation, large enough to explain the Pioneer anomaly, is in
%obvious contradiction with the planetary ephemerides. This becomes
%particularly clear if one considers the orbit of Neptune. [...]
%The influence of an additional radial acceleration of $8\times
%10^{-10}$  m/s$^2$ on Neptune is conveniently parameterized in a
%change of effective reduced Solar mass, $\mu_{\odot}$ felt by the
%planet \cite{redmass}. The resulting value,
%$\Delta\mu_{\odot}=1.4\times 10^{-4}\mu_{\odot}$, is nearly two
%orders of magnitude beyond the current observational constraint of
%$\Delta\mu_{\odot}=-1.9\pm 1.8\times 10^{-6}\mu_{\odot}$.
%Similarly, Pioneer 11 data contradict the Uranus ephemerides by
%more than one order of magnitude.''
Another recent paper in which
a comparison between planetary observations and a theoretically
derived quantity related to the Pioneer anomaly is attempted
yielding similar conclusions is \ct{tangen}.

In our paper we first work out, both analytically and numerically,
the orbital perturbations induced by a disturbing acceleration
like the Pioneer one on the Keplerian orbital elements of a
planet. We find that the semimajor axis, the eccentricity, the
perihelion and the mean anomaly of a test particle are affected by
short-periods signatures and the perihelion and the mean anomaly
undergo also long-period, secular effects. Then, numerical
simulations of the temporal evolutions of the directly observable
right ascensions and declinations of the planets induced by a
Pioneer-like acceleration are performed. Contrary to the position
of the authors of \cite{Page06} and in according to the
conclusions of \cite{Izzo06, tangen}, it turns out that the
predicted anomalous effects of this kind would be sufficiently
large to be well detected with the present-day level of accuracy
in planetary orbit determination \cite{Pit05a}, even by
considering errors considerably larger than those released. Their
absence tell us that in the outer region of the Solar System
within 20-40 AU there is no any unknown force field, with the
characteristics of that experienced by the Pioneer spacecraft,
which affects the motion of the major planets.

\section{The Gauss rate equations}

The Gauss equations for the variations of the semimajor axis $a$,
the eccentricity $e$, the inclination $i$, the longitude of the
ascending node $\Omega$, the argument of pericentre $\omega$ and
the mean anomaly $\mathcal{M}$ of a test particle of mass $m$ in
the gravitational field of a body $M$ can be derived quite
generally from \cite{mil}\eqi
\ddot{\boldsymbol{r}}+\rp{\mu}{r^3}\boldsymbol{r}=\boldsymbol{A},\label{twobody}\eqf
where $\mu=G(M+m)$ and $\boldsymbol{r}=\boldsymbol{r}_m
-\boldsymbol{r}_M$. \Rfr{twobody} holds for every perturbing
acceleration $\boldsymbol{A}$, whatever its cause or size.
 The variations of the elements are, for an entirely radial
 acceleration
%
%\bar
%\dert{a}{t} & = & \rp{2}{n\sqrt{1-e^2}}\left[eA_r\sin f+A_t\left(\rp{p}{r}\right)\right],\lb{smax}\\
%\dert{e}{t} & = & \rp{\sqrt{1-e^2}}{na}\left\{A_r\sin f+A_t\left[\cos f+\rp{1}{e}\left(1-\rp{r}{a}\right)\right]\right\},\\
%\dert{i}{t} & = & \rp{1}{na\sqrt{1-e^2}}\ A_n\left(\rp{r}{a}\right)\cos (\omega+f),\lb{in}\\
%\dert{\Omega}{t} & = & \rp{1}{na\sin i\sqrt{1-e^2}}\ A_n\left(\rp{r}{a}\right)\sin (\omega+f),\lb{nod}\\
%\dert{\omega}{t} & = & -\cos i\dert{\Omega}{t}+\rp{\sqrt{1-e^2}}{nae}\left[-A_r\cos f+A_t\left(1+\rp{r}{p}\right)\sin f\right],\lb{perigeo}\\
%\dert{\mathcal{M}}{t} & = & n -\rp{2}{na}\
%A_r\left(\rp{r}{a}\right)-\sqrt{1-e^2}\left(\dert{\omega}{t}+\cos
%i\dert{\Omega}{t}\right),\lb{manom} \ear
%
%
%
%
%If the perturbing acceleration is entirely radial
%\rfrs{smax}{manom} reduce to
%
%

\bar
\dert{a}{t} & = & \rp{2e}{n\sqrt{1-e^2}}\ A_r\sin f,\lb{smax2}\\
\dert{e}{t} & = & \rp{\sqrt{1-e^2}}{na}\ A_r\sin f,\\
\dert{i}{t} & = & 0,\lb{in2}\\
\dert{\Omega}{t} & = & 0,\lb{nod2}\\
\dert{\omega}{t} & = & -\rp{\sqrt{1-e^2}}{nae}\ A_r\cos f,\lb{perigeo2}\\
\dert{\mathcal{M}}{t} & = & n -\rp{2}{na}\
A_r\left(\rp{r}{a}\right)-\sqrt{1-e^2}\
\dert{\omega}{t},\lb{manom2} \ear
in which $n=2\pi/P$ is the  mean motion, $P$ is the test
particle's orbital period, $f$ is the true anomaly counted from
the pericentre, $p=a(1-e^2)$ is the semilactus rectum of the
Keplerian ellipse and $A_r$ is the  projection of the perturbing
acceleration $\boldsymbol{A}$ on the radial direction of the
co-moving frame
$\{\boldsymbol{\hat{r}},\boldsymbol{\hat{t}},\boldsymbol{\hat{n}}\}$.
As can be noted from \rfrs{in2}{nod2}, the inclination and the
node are not perturbed by such a perturbing acceleration.

\section{The orbital effects induced by a constant radial acceleration}
\subsection{Analytical calculation}\lb{analy} The ratio of the
Pioneer anomalous acceleration to the Newtonian monopole term
ranges from $5\times 10^{-5}$ to $2\times 10^{-4}$ in the region
20-40 AU, so that \rfrs{smax2}{manom2} can be treated in the
perturbative way. A treatment of the effects of a constant radial
acceleration of arbitrary magnitude in the two-body problem can be
found in \cite{boltz, pruss}.

The right-hand-sides of  \rfrs{smax2}{manom2} have to be evaluated
on the unperturbed Keplerian ellipse.
%which, in terms of, e.g.,
%$f$, is given by \eqi r=\rp{a(1-e^2)}{1+e\cos f }.\lb{kep}\eqf
The
secular effects can be obtained by averaging over one orbital
period the right-hand-sides of \rfrs{smax2}{manom2}.

The integration of \rfrs{smax2}{manom2} is easier if the eccentric
anomaly $E$, defined as $\mathcal{M}=E-e\sin E$, is used instead
of the true anomaly $f$. The most important relations in terms of
$E$ are \bar dt &=&
\rp{(1-e\cos E)}{n}dE,\lb{ecc1}\\
r &=& a(1-e\cos E),\\
\cos f &=& \rp{(\cos E-e)}{1-e\cos E},\\
\sin f &=& \rp{\sqrt{1-e^2}\sin E}{1-e\cos E}.\lb{ecc4}\ear

By using \rfrs{ecc1}{ecc4} the shifts of the Keplerian orbital
elements over a time span shorter than one orbital period, i.e.
$E-E_0< 2\pi$, become

\bar \Delta a &=& -\left.\frac{2 e A_r}{n^2}\cos \xi\right|_{E_0}^E,\lb{smax3}\\
\Delta e &=& -\left.\frac{(1-e^2)A_r}{n^2 a}\cos \xi\right|_{E_0}^E,\lb{ecce3}\\
%\Delta i &=&0,\\
%\Delta\Omega &=& 0,\\
\Delta\omega &=& \left.\frac {
\sqrt{1-e^2} A_r }{n^2 a e} (e
\xi-\sin \xi)\right|_{E_0}^{E}, \lb{perigeo3}\\
\Delta {\mathcal{M}} &=& n\Delta t+\left.\rp{A_r}{n^2
a}\left\{-3\xi -e^2\rp{\sin 2
\xi}{2}+\left[4e+\rp{(1-e^2)}{e}\right]\sin\xi\right\}\right|_{E_0}^{E}.\lb{manom3}
\ear

%
%\bar
%\Delta{a} & = & \left.\frac{2 A_r (1-e^2)}{n^2}\left(\frac{1}{1+e\cos \xi}\right)\right|_{f_0}^{f},\lb{smax3}\\
%\Delta{e} & = & \left.\frac{ A_r (1-e^2)^{5/2}  }{a e n^2}\left(\frac{1}{1+e\cos \xi}\right)\right|_{f_0}^{f},\lb{ecce3}\\
%\Delta{i} & = & 0,\lb{in3}\\
%\Delta{\Omega} & = & 0,\lb{nod3}\\
%\Delta{\omega} & = &
%\frac{A_r(1-e^2)^{2}}{ean^2}\left[\rp{2e}{(1-e^2)^{3/2}}\arctan\left(\sqrt{\frac{1-e}{1+e}}
%\tan\frac{\xi}{2}\right)-\right.\nonumber\\
%&-&\left.\left.\rp{\sin \xi}{(1-e^2)(1+e\cos \xi)}\right]\right|_{f_0}^{f},\lb{perigeo3}\\
%\Delta{\mathcal{M}} & = & n\Delta t -\rp{2A_r (1-e^2)^{5/2}}{n^2
%a}\left[
%\rp{2+e^2}{(1-e^2)^{5/2}}\arctan\left(\sqrt{\frac{1-e}{1+e}}
%\tan\frac{\xi}{2}\right)+\right.\nonumber\\
%&+&\left.\left.\rp{e(-4+e^2-3e\cos\xi)\sin\xi}{2(1-e^2)^2(1+e\cos\xi)^2}\right]\right|_{f_0}^{f}
%-\sqrt{1-e^2}\ \Delta\omega,\lb{manom3} \ear
%

From \rfrs{smax3}{ecce3} it can straightforwardly be noted that
the shifts over one full orbital revolution, i.e. from $E_0$ to
$E_0+2\pi$, vanish for the semimajor axis and the eccentricity, so
that  there are no net secular orbital effects for such Keplerian
orbital elements. It is not so for the pericentre and the mean
anomaly whose secular rates are, from \rfrs{perigeo3}{manom3} \bar
\left\langle\dert\omega t\right\rangle & = & \frac{A_r\sqrt{1-e^2}}{na},\lb{secperi}\\
\left\langle\dert{\mathcal{M}}t\right\rangle & = &
-\frac{3A_r}{na}.\lb{secmanom} \ear

The planets of the Solar System show moderate eccentricities (and
inclinations). A widely used orbital element for such kind of
orbits is the mean longitude $\lambda=\Omega+\omega+\mathcal{M}$.
From \rfrs{perigeo3}{manom3} we have, to order $\mathcal{O}(e^2)$

\eqi\Delta\lambda\sim n\Delta t
-\left.\rp{2A_r}{n^2a}\left[\left(1+\rp{e^2}{4}\right)\xi+\rp{e^2}{4}\sin
2\xi -\rp{7}{4}e\sin\xi\right]\right|_{E_0}^E.\lb{approan}\eqf The
secular rate is, thus
\eqi\left\langle\frac{d\lambda}{dt}\right\rangle\sim n-\rp{
2A_r}{na}\left(1+\rp{e^2}{4}\right).\eqf
\subsection{Application to Uranus, Neptune and Pluto}
Since the Pioneer anomaly began to manifest itself at the 20 AU
threshold, we will apply such analytical results  to Uranus,
Neptune and Pluto whose semimajor axes are about 19 AU, 30 AU and
40 AU, respectively.

The secular effects which would be induced on the perihelia and
the mean longitudes of Uranus, Neptune and Pluto by a constant
radial acceleration of the same magnitude as that acting upon
Pioneer are listed in Table \ref{secrate}. They amount to tens and
hundreds of arcseconds/century (\asec). In regard to their
detection, it must be noted that the orbital periods of the
investigated planets, which are 84 years for Uranus, 164 years for
Neptune and 248 years for Pluto, are comparable or larger than the
time span in which accurate observations were collected. Thus, it
is not yet possible to single out orbital effects averaged over
one full orbital revolution for Neptune and Pluto.

In Table \ref{Pit}, which reproduces Table 4 obtained by E.V.
Pitjeva in \cite{Pit05a} by processing almost one century of data
of various kinds with the latest EPM2004 ephemerides released by
the Institute of Applied Astronomy of Russian Academy of Sciences,
the formal standard deviations of the orbital elements of the nine
major planets of the Solar System are reported. The optical
observations used for Uranus, Neptune and Pluto are 23612  and
cover 90 years from 1913 (1914 for Pluto) to 2003 (see Table 3 of
\cite{Pit05a}). It turns out from Table \ref{Pit} that the
accuracy in determining the perihelia and mean longitudes of
Uranus, Neptune and Pluto would be largely adequate to see so huge
anomalous signatures, even if we consider that realistic errors
could be ten-thirty times larger. Indeed, from the formal results
of Table \ref{Pit} we have uncertainties of $0.1 ''$, $2 ''$ and
$0.3 ''$ for the perihelia of Uranus, Neptune and Pluto,
respectively; if we multiply them by a factor 30 we get $3 ''$,
$60 ''$ and $9 ''$, respectively while the anomalous perihelion
shifts over 90 years would amount to $75.1 ''$, $93 ''$ and $104.6
''$, respectively. In the case of the mean longitudes, the formal
uncertainties amount to $0.008 ''$, $0.035 ''$ and $0.079 ''$,
respectively which become $0.2''$, $1 ''$ and $2''$  if rescaled
by a factor 30; the anomalous shifts on $\lambda$ would be
$150.6''$, $188.5''$ and $217.7''$, respectively. Note also that
the systematic errors in the Keplerian mean motions $\delta
n=(3/2)\sqrt{GM/a^5}\delta a$ due to the (formal) uncertainties in
the semimajor axis $a$ amount to $0.03$ \asec, $0.1$\asec\ and
$0.4$\asec\ only: even re-scaled by a factor 30, they remain well
smaller than the anomalous shifts. The impact of the mismodelling
in solar $GM$, assumed $\delta(GM)=8\times 10^9$ m$^3$ s$^{-2}$
(http://ssd.jpl.nasa.gov/astro$\_$constants.html), on the mean
longitudes is even smaller.

\begin{figure}[htbp]
\begin{center}
\includegraphics[width=14cm,height=12cm,angle=0]{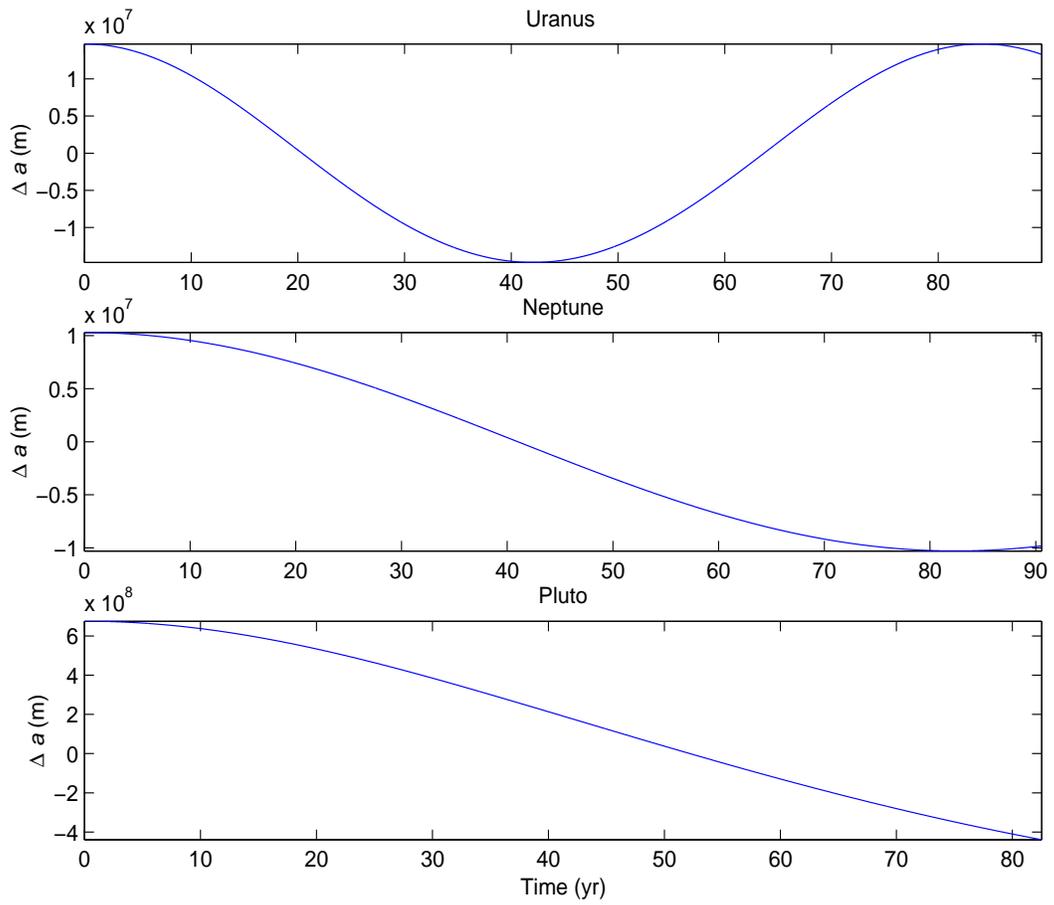}
\end{center}
\caption{\label{Pio_da} Semimajor axis shifts $\Delta a$, in m,
for Uranus ($a=19.19$ AU), Neptune ($a=30.06$ AU) and Pluto
($a=39.48$ AU) over $T=90$ years.}
\end{figure}

\begin{figure}[htbp]
\begin{center}
\includegraphics[width=14cm,height=12cm,angle=0]{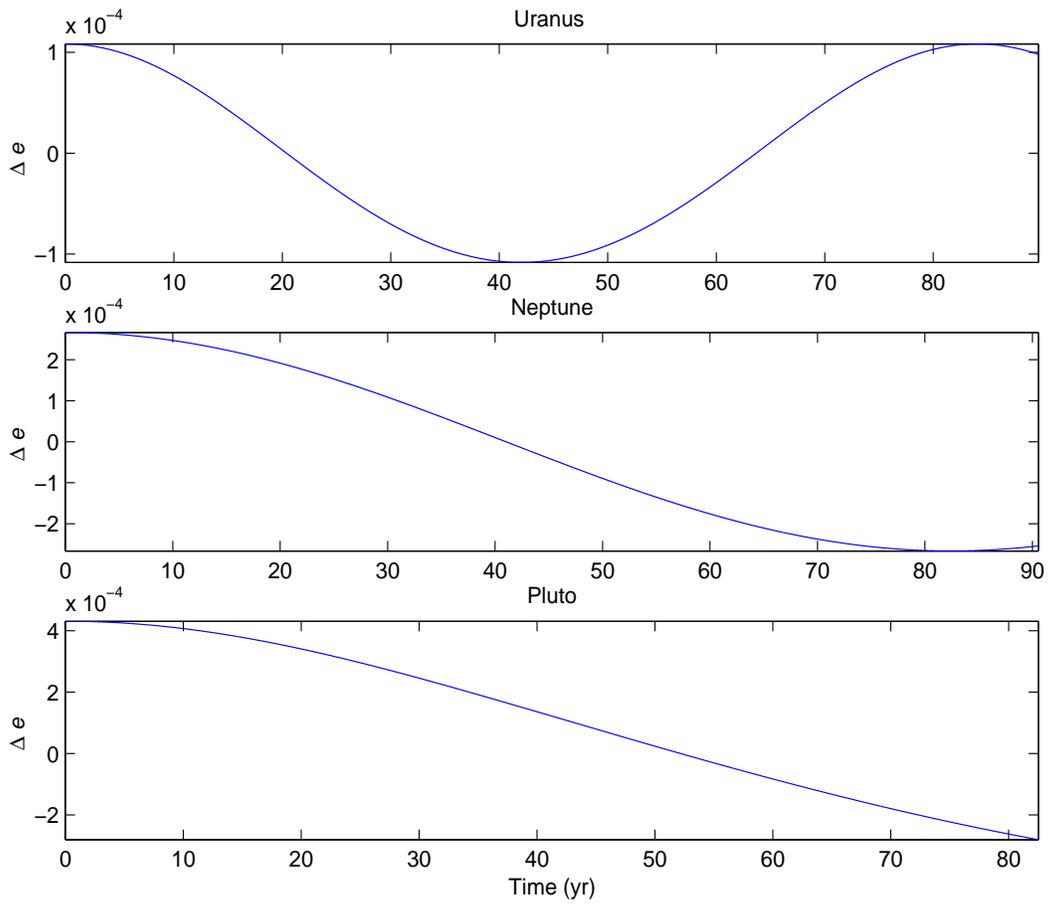}
\end{center}
\caption{\label{Pio_de} Eccentricity shifts $\Delta e$ for Uranus
($a=19.19$ AU), Neptune ($a=30.06$ AU) and Pluto ($a=39.48$ AU)
over $T=90$ years.}
\end{figure}

\begin{figure}[htbp]
\begin{center}
\includegraphics[width=14cm,height=12cm,angle=0]{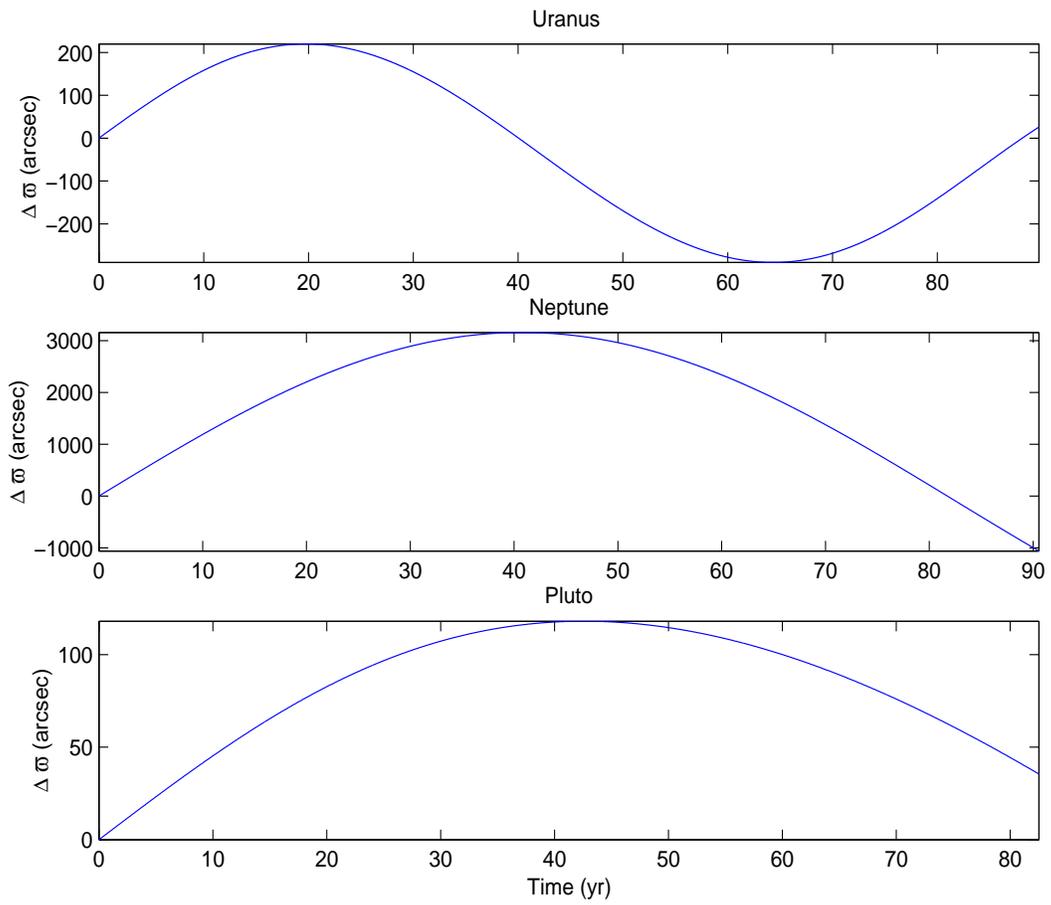}
\end{center}
\caption{\label{Pio_do} Longitude of perihelion shifts $\Delta
\varpi$, in $''$, for Uranus ($a=19.19$ AU), Neptune ($a=30.06$
AU) and Pluto ($a=39.48$ AU) over $T=90$ years.}
\end{figure}

\begin{figure}[htbp]
\begin{center}
\includegraphics[width=14cm,height=12cm,angle=0]{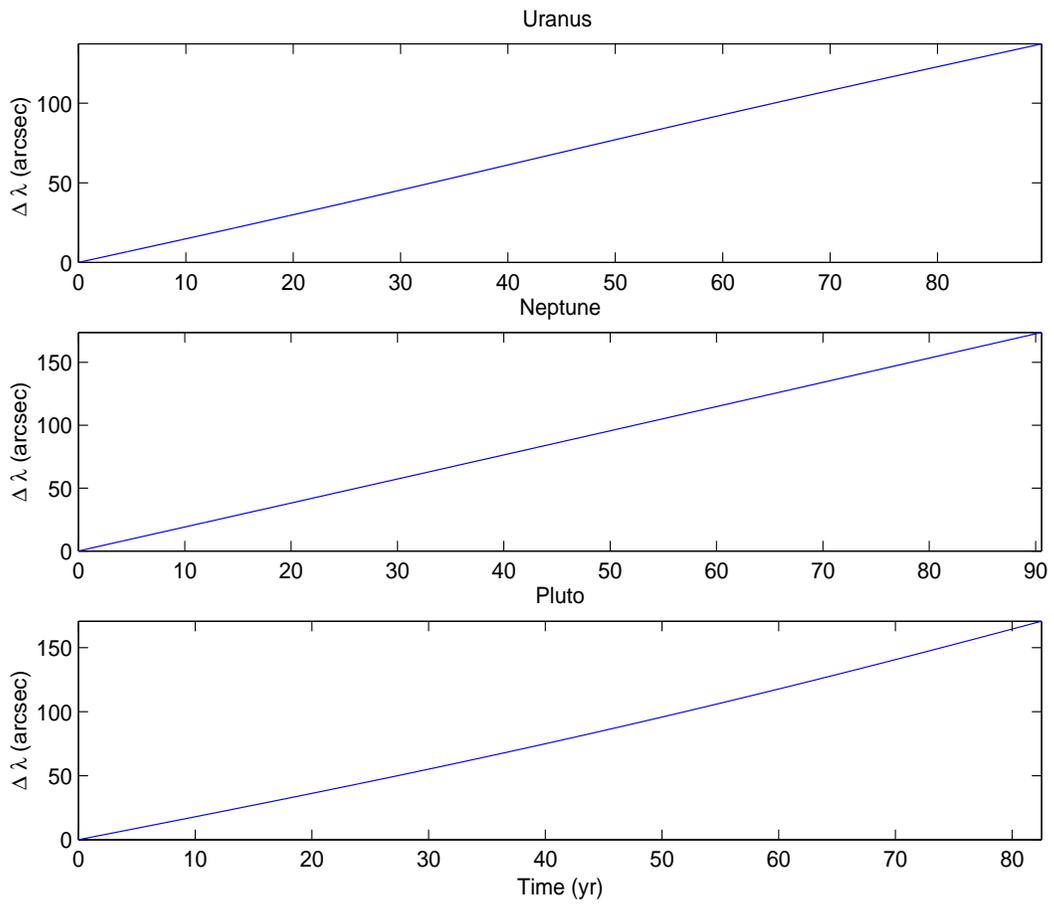}
\end{center}
\caption{\label{Pio_dl} Mean longitude  shifts $\Delta \lambda$,
in $''$, for Uranus ($a=19.19$ AU), Neptune ($a=30.06$ AU) and
Pluto ($a=39.48$ AU) over $T=90$ years. }
\end{figure}

The very long orbital periods of Uranus, Neptune and Pluto make
also the short-period signals of interest, in principle, because
they would resemble polynomial signals over observational time
spans like those currently available. This could be useful
especially for the semimajor axis and the eccentricity which do
not exhibit secular rates. In Figure \ref{Pio_da}, Figure
\ref{Pio_de}, Figure \ref{Pio_do} and \ref{Pio_dl} we plot the
total, i.e. linear (when present) and sinusoidal, integrated
shifts of \rfrs{smax3}{perigeo3} and \rfr{approan} for the
semimajor axis, the eccentricity, the perihelion and the mean
longitude, respectively over $90$ years. In Table \ref{ampli} we
quote the amplitudes ${\mathcal{A}}$ of the short-period signals
for Uranus, Neptune and Pluto. Anomalous signatures of such size
would be detectable as well, but nothing like that is mentioned in
Pitjeva's work.

\section{The orbital effects induced by a constant radial acceleration: numerical
simulations and comparison with data}\lb{nume}

The Keplerian orbital elements are not directly measured
quantities: they are related to data in an indirect way. For the
outer planets the true observables are the right ascension
$\alpha$ and the declination $\delta$. In Figure 2 of
\cite{Pit05a} there are the observational residuals of
$\alpha\cos\delta$ and $\delta$ for Uranus, Neptune and Pluto from
1913 to 2004. In order to make a direct comparison with them, in
this Section we numerically compute theoretical Pioneer-No Pioneer
(P-NP) residuals for the same quantities over the same time span.
%%%%%%  Giudice contribution

\begin{figure}[htbp]
\begin{center}
\includegraphics[width=14cm,height=12cm,angle=0]{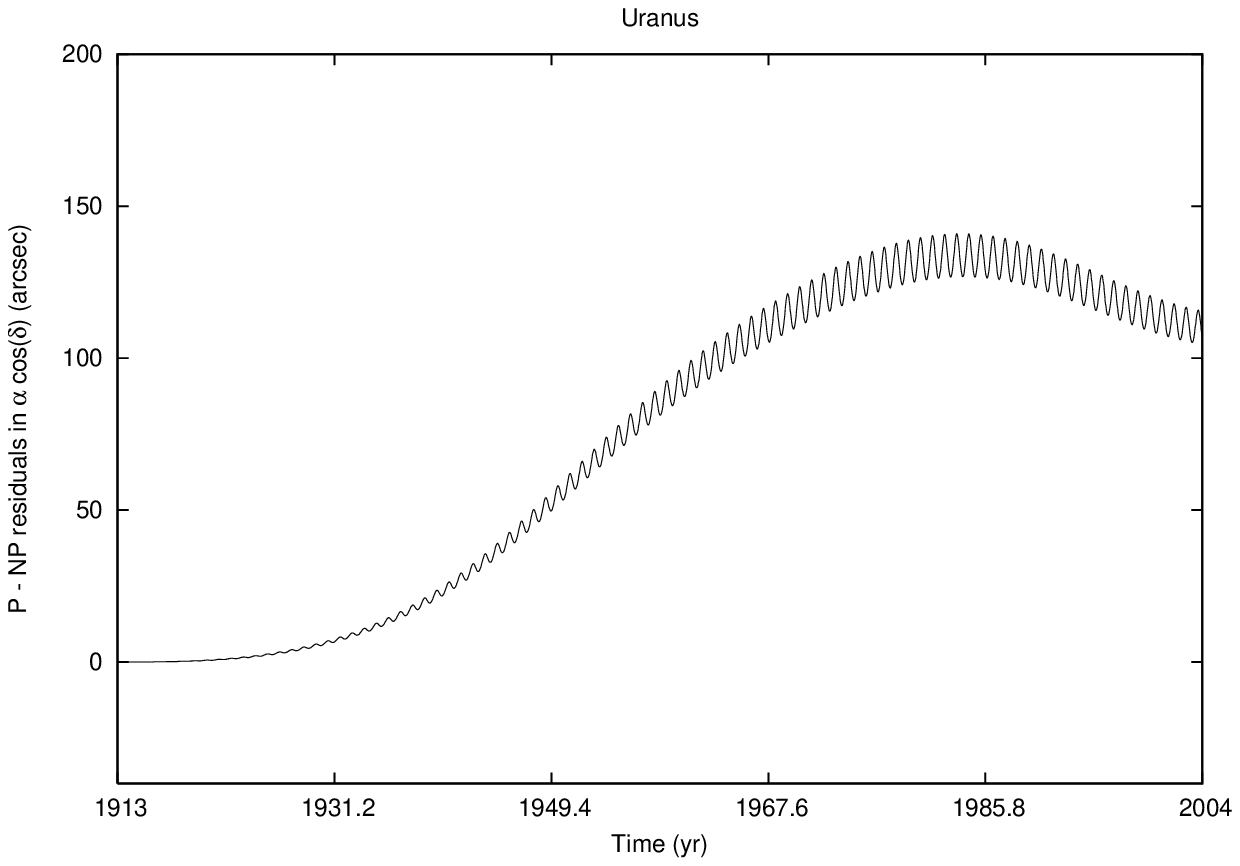}
\end{center}
\caption{\label{ur_ra} Numerically propagated P-NP residuals for
$\alpha\cos\delta$ of Uranus from 1913 to 2004. Compare it with
the third panel from the bottom of Figure 2 of \ct{Pit05a}.}
\end{figure}

\begin{figure}[htbp]
\begin{center}
\includegraphics[width=14cm,height=12cm,angle=0]{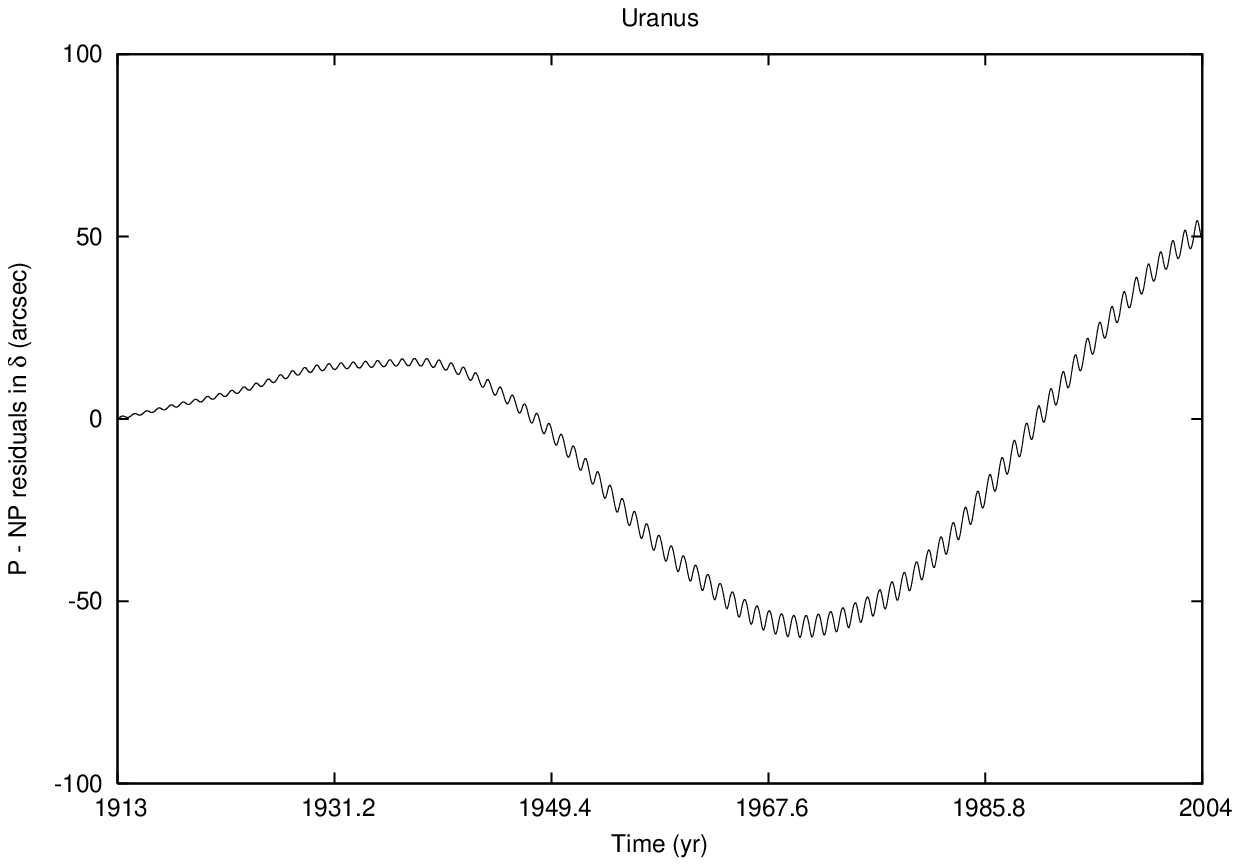}
\end{center}
\caption{\label{ur_dec} Numerically propagated P-NP residuals for
$\delta$ of Uranus from 1913 to 2004. Compare it with the third
panel from the bottom of Figure 2 of \ct{Pit05a}.}
\end{figure}

\begin{figure}[htbp]
\begin{center}
\includegraphics[width=14cm,height=12cm,angle=0]{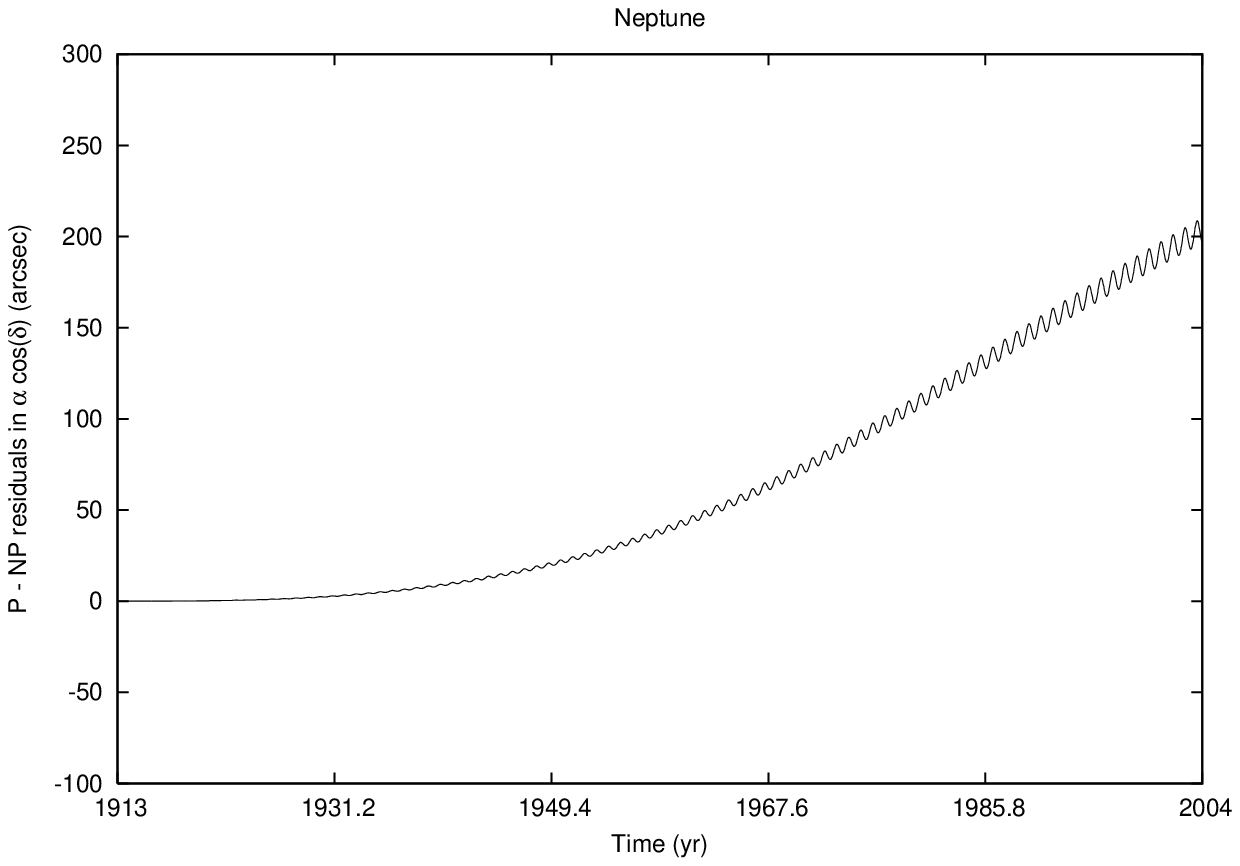}
\end{center}
\caption{\label{ne_ra} Numerically propagated P-NP residuals for
$\alpha\cos\delta$ of Neptune from 1913 to 2004. Compare it with
the second panel from the bottom of Figure 2 of \ct{Pit05a}.}
\end{figure}

\begin{figure}[htbp]
\begin{center}
\includegraphics[width=14cm,height=12cm,angle=0]{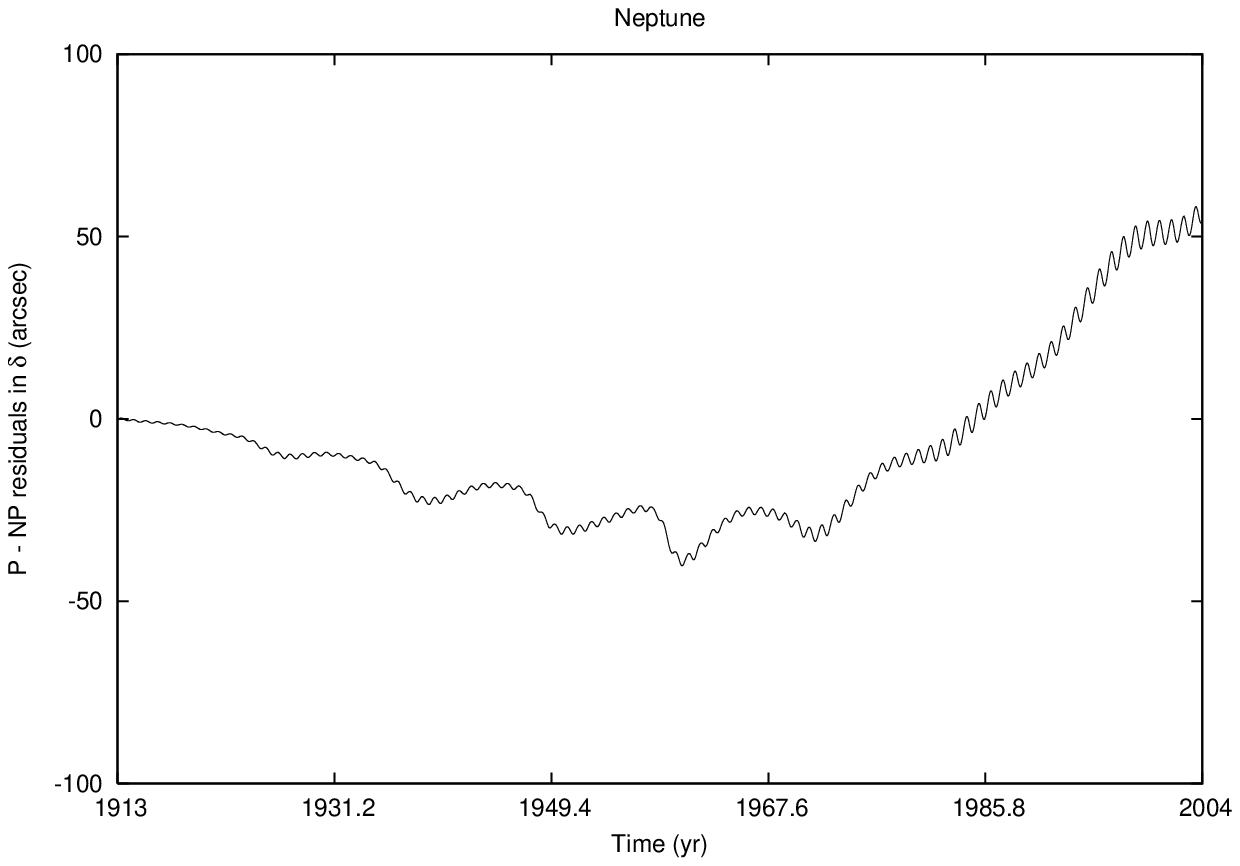}
\end{center}
\caption{\label{ne_dec} Numerically propagated P-NP residuals for
$\delta$ of Neptune from 1913 to 2004. Compare it with the second
panel from the bottom of Figure 2 of \ct{Pit05a}.}
\end{figure}

\begin{figure}[htbp]
\begin{center}
\includegraphics[width=14cm,height=12cm,angle=0]{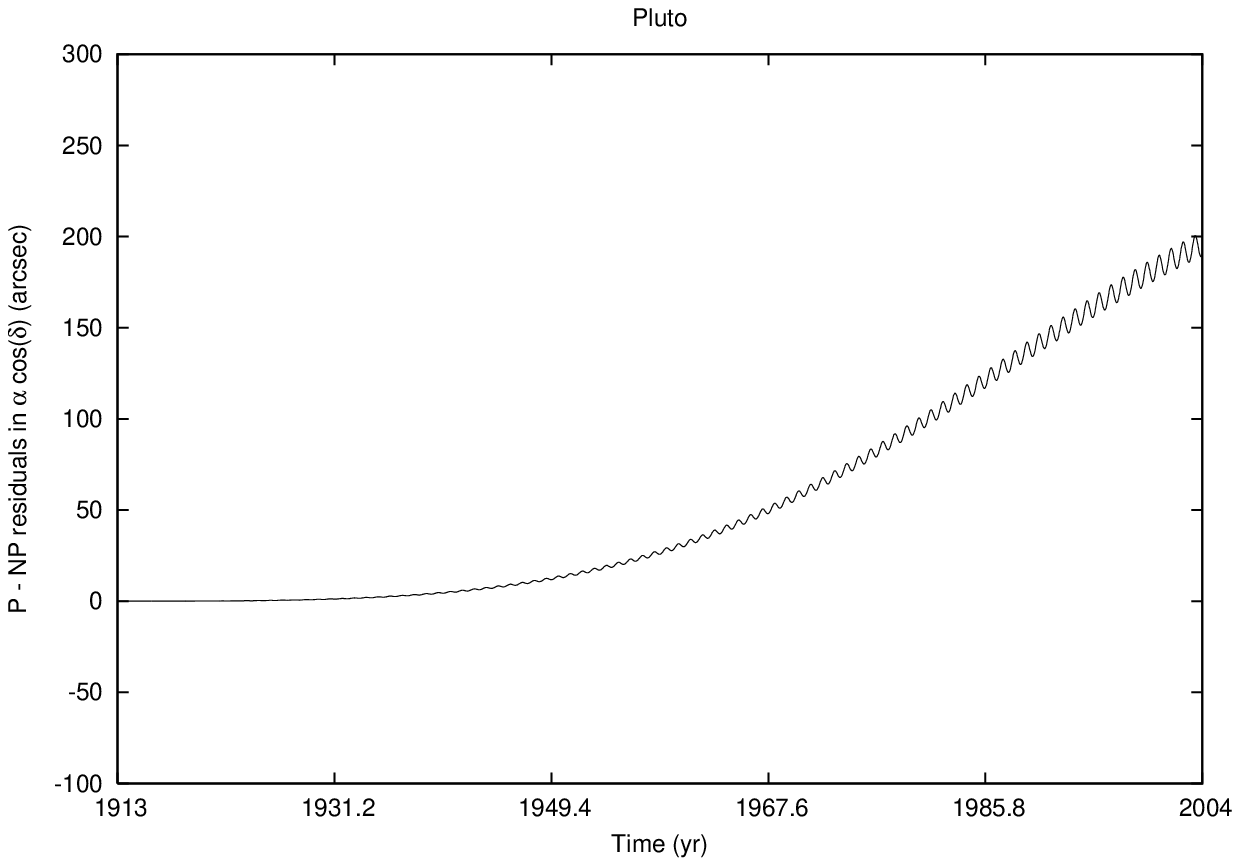}
\end{center}
\caption{\label{pl_ra} Numerically propagated P-NP residuals for
$\alpha\cos\delta$ of Pluto from 1913 to 2004. Compare it with the
first panel from the bottom of Figure 2 of \ct{Pit05a}.}
\end{figure}

\begin{figure}[htbp]
\begin{center}
\includegraphics[width=14cm,height=12cm,angle=0]{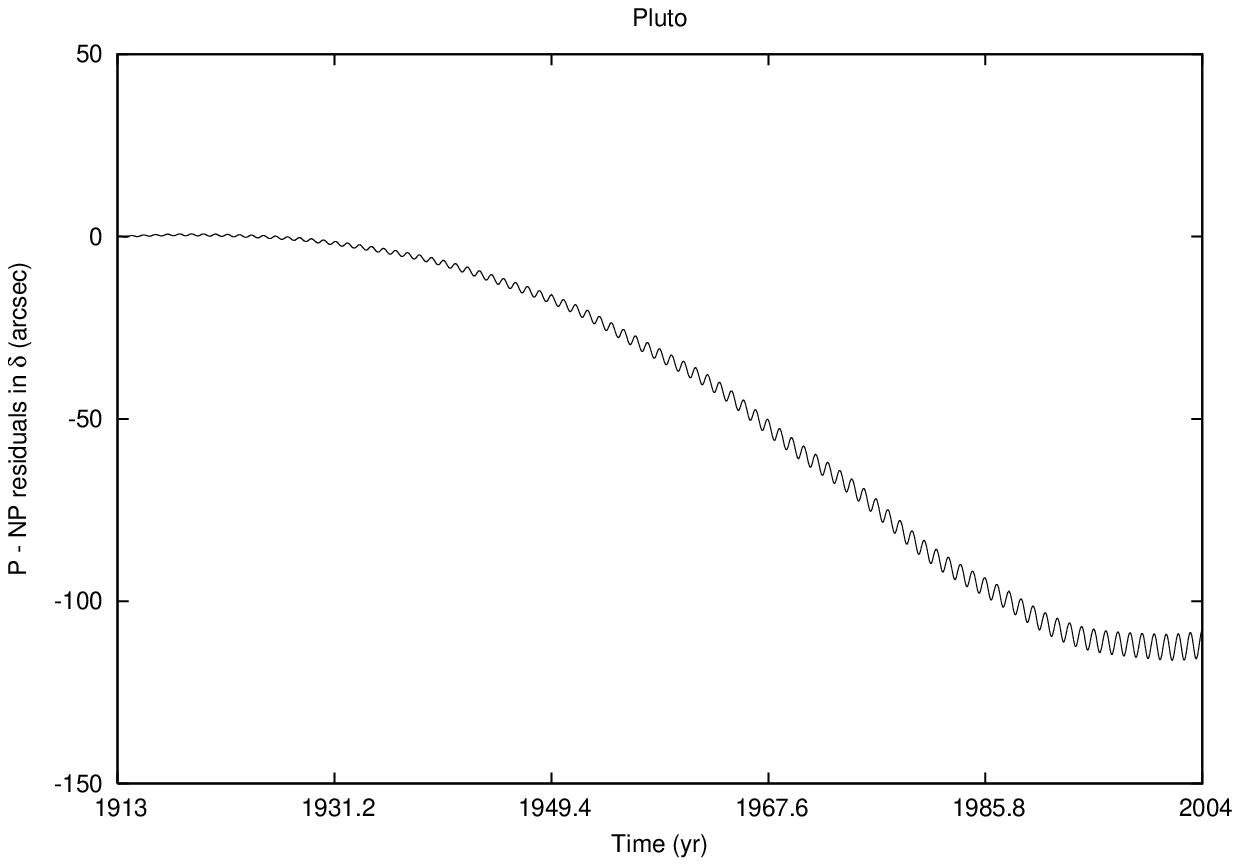}
\end{center}
\caption{\label{pl_dec} Numerically propagated P-NP residuals for
$\delta$ of Pluto from 1913 to 2004. Compare it with the first
panel from the bottom of Figure 2 of \ct{Pit05a}.}
\end{figure}

The first step of this process consists in a direct integration of
the gravitational equations in Cartesian rectangular coordinates.
%We remark here that the numerical integration of ordinary
%differential equations is, in a sense, an art, because the general
%purpose methods, as the explicit Runge-Kutta one, have to be
%disregarded when the problem belongs to particular classes; this
%is the case of the long-term (and also medium term) integration of
%the Solar System, because this is a conservative problem, and the
%Runge-Kutta explicit method tend to variate the energy of the
%system, not because of the intrinsic physical characteristics  of
%problem, but because of the intrinsic mathematical characteristics
%of the method. For this reason other methods have been developed
%and employed, and in particular the symplectic ones. Between them
%there are the the implicit Runge-Kutta methods, with which is
%related the Everhart-Radau method \cite{Everhart74, Everhart85}.
%The Bulirsch-Stoer method \cite{StoBul} is also widely used. In
%every case it is mandatory that the employed method is of a
%sufficiently elevated order, so to make possible the use of an as
%longest step as possible, so to minimize the rounding-off errors.
A number of software packages  to perform this goal are freely
available on the Internet; between them there are OrbFit ({\tt
http://newton.dm.unipi.it/orbfit/}), ORSA (Orbit Reconstruction,
Simulation and Analysis, {\tt http://orsa.sourceforge.net/}, by P.
Tricarico), and Mercury \cite{Chamb99}. The last one has been
chosen because it has the simplest use and allows to introduce
without difficulty user-defined perturbing forces, such the
Pioneer one. It is introduced through the definition of the
corresponding acceleration, whose components in $x,y$ and $z$ in
an heliocentric frame are given. In this implementation this
acceleration has been considered only for heliocentric distances
greater then 15.0 AU, so to include the whole orbit of Uranus,
totally excluding Saturn. This package uses the following
integration methods
\begin{itemize}
  \item second order mixed-variable symplectic
  \item Bulirsh-Stoer (general)
  \item Bulirsh-Stoer (conservative systems)
  \item Radau 15th order
  \item hybrid (symplectic / Bulirsh-Stoer)
\end{itemize}
Since the pros and cons of a method depend to a certain extent on
the specific problem to be solved, a number of tests have been
performed, using most of the proposed methods and various initial
conditions in order to choose the most suitable approach to our
problem.
%
%
%The output of ORSA has been used as benchmark result: as a matter
%of fact, this package ties its output to JPL ephemerides, so
%ensuring a better precision. The integration with ORSA has been
performed form JD 2410000.5 to JD 2460000.5, for the nine planets
As a result, the 15th order Everhart-Radau algorithm has been
chosen for the Pioneer problem and initial conditions at JD
2410000.5. The output is in $XYZ$ barycentric, with both
integration and tabulation steps of 10.0 days. The $XYZ$
coordinates of  the Earth's center are obtained from ORSA. The
right ascensions and declinations of the three planets have been
obtained in the standard way (see e.g. \cite{Meeus}), and the
residuals in $\alpha\cos\delta$ and $\delta$ have been plotted in
Figures \ref{ur_ra}-\ref{pl_dec}.
%
%
%
%%%%%% End of Giudice contribution
%
As already noted, Figure 2 of \cite{Pit05a} shows the
observational residuals, in $''$, of $\alpha\cos\delta$ and
$\delta$ for Jupiter, Saturn, Uranus, Neptune and Pluto; the scale
is $\pm 5$ $''$ and no secular or semi-secular trends can be
recognized by visual inspection. Our numerically produced P-NP
residuals of Figures \ref{ur_ra}-\ref{pl_dec} can be
straightforwardly compared to the last three panels (from the
bottom) of Figure 2 of \cite{Pit05a} yielding a direct and
unambiguous confrontation with the observations. As can be noted,
the pattern which would be induced by a Pioneer-like acceleration
on the planetary motions is absent in the data.

\section{Summary and conclusions}
In this paper we have considered the orbital motions of Uranus,
Neptune and Pluto to investigate, from a purely phenomenological
point of view, if an anomalous force field inducing a constant,
uniform Sunward acceleration of $8.74\times 10^{-10}$ m s$^{-2}$,
as that experienced by the Pioneer spacecraft, affects their
motions which occur  in the 20-40 AU region of the Solar System,
well within the range in which the Pioneer data show the anomalous
signature. We have worked out both analytically and numerically
the perturbations induced by such a disturbing acceleration on the
Keplerian orbital elements and on the directly observable right
ascensions and declinations of Uranus, Neptune and Pluto. It turns
out that there are long-period, secular rates on the perihelion
and the mean anomaly and short-period effects on the semimajor
axis, the eccentricity, the perihelion and the mean anomaly which
map onto certain particular patterns of the right ascensions and
declinations. Such anomalous signatures would be large enough to
be detected according to the present-day accuracy in orbit
determination, but there is no trace of them in the currently
available data. This result restricts the possible causes of the
Pioneer anomaly to some unknown forces which violate the
equivalence principle in a very strange way or to some
non-gravitational mechanisms peculiar to the spacecraft.

\section*{Acknowledgements}
L.I. gratefully thanks G. L. Page, D. Izzo and J. Katz for having
kindly submitted to his attention their useful papers. Thanks also
to S. Turyshev  for interesting discussions.

%-----------------------------------------

\newpage

\begin{table}
\caption{Secular effects, in arcseconds/century (\asec), induced
by a constant radial acceleration of magnitude $8.74\times
10^{-10}$ m s$^{-2}$ on the perihelion and mean longitude of
Uranus, Neptune and Pluto. }\label{secrate}
\begin{tabular}{@{\hspace{0pt}}lll}
\hline\noalign{\smallskip} Planet & $\varpi$ & $\lambda$\\
\noalign{\smallskip}\hline\noalign{\smallskip}
Uranus &  -83.5 & 167.401\\
Neptune & -104 & 209.48\\
Pluto & -116.2 & 241.89\\
 \noalign{\smallskip}\hline
\end{tabular}
\end{table}
\begin{table}
\caption{Formal standard statistical errors in the non-singular
planetary orbital elements, from Table 4 of \cite{Pit05a}. The
latest EPM2004 ephemerides have been used. The realistic errors
may be up to one order of magnitude larger. The units for the
angular parameters are milliarcseconds (mas).}\label{Pit}
\begin{tabular}{@{\hspace{0pt}}lllllll}
\hline\noalign{\smallskip} Planet & $a$ (m) & $\sin i\cos\Omega$
(mas)& $\sin i\sin\Omega$ (mas)& $e\cos\varpi$ (mas)&
$e\sin\varpi$ (mas) & $\lambda$ (mas)
\\
\noalign{\smallskip}\hline\noalign{\smallskip}
Mercury & 0.105 & 1.654 & 1.525 & 0.123 & 0.099 & 0.375\\
Venus & 0.329 & 0.567 & 0.567 & 0.041 &  0.043 & 0.187\\
Earth & 0.146 &- & -& 0.001 &  0.001 &  -\\
Mars & 0.657 &0.003 & 0.004 & 0.001 & 0.001 & 0.003\\
Jupiter & 639 & 2.410 & 2.207 & 1.280 &  1.170 & 1.109\\
Saturn & 4222 & 3.237 &  4.085 & 3.858 & 2.975 &  3.474 \\
Uranus & 38484 & 4.072 & 6.143 & 4.896 & 3.361 & 8.818\\
Neptune & 478532 & 4.214 & 8.600 & 14.066 & 18.687 & 35.163\\
Pluto & 3463309 & 6.899 & 14.940 &  82.888 & 36.700 & 79.089\\
 \noalign{\smallskip}\hline
\end{tabular}
\end{table}
\begin{table}
\caption{ Amplitudes ${\mathcal{A}}$ of the integrated
short-period effects induced by a constant radial acceleration
with magnitude $8.74\times 10^{-10}$ m s$^{-2}$ on the orbital
elements of Uranus, Neptune and Pluto. Note that the values of ${
\mathcal{A} }_{\rm \Delta\lambda}$ and the present-day errors in
$\lambda$ fully justify the approximation used to work out
\rfr{approan}. }\label{ampli}
\begin{tabular}{@{\hspace{0pt}}lllll}
\hline\noalign{\smallskip} Planet & ${\mathcal{A}}_{\Delta a}$ (m)
& ${\mathcal{A}}_{\Delta e}$ &
${\mathcal{A}}_{\Delta\varpi}$ ($''$) & ${\mathcal{A}}_{\Delta\lambda}$ ($''$)\\
\noalign{\smallskip}\hline\noalign{\smallskip}
Uranus & $1.470\times 10^{7}$  & $1\times 10^{-4}$ & 237.1 & $1.848$\\
Neptune & $1.03\times 10^7$ & $2.6\times 10^{-4}$  & 3201 & $0.82$\\
Pluto & $6.7\times 10^8$ & $4\times 10^{-4}$ & $184.5$ & $41.64$ \\
 \noalign{\smallskip}\hline
\end{tabular}
\end{table}

\newpage

%
%-------------------------------------------------
\end{document}